\documentstyle[sprocl]{article}

\def\be{\begin{equation}}
\def\ee{\end{equation}}
\def\bea{\begin{eqnarray}}
\def\eea{\end{eqnarray}}

\def\IR{{\hbox{{\rm I}\kern-.2em\hbox{\rm R}}}}
\def\IH{{\hbox{{\rm I}\kern-.2em\hbox{\rm H}}}}
\def\IC{{\ \hbox{{\rm I}\kern-.6em\hbox{\bf C}}}}
\def\IZ{{\hbox{{\rm Z}\kern-.4em\hbox{\rm Z}}}}

\begin{document}

\begin{flushright}
\begin{small}
hep-th/9806071\\
UPR/808-T \\
June 1998 \\
\end{small}
\end{flushright}

\title{Anti-DeSitter Spaces and Nonextreme Black Holes
\footnote{To appear in the proceedings of PASCOS '98. The actual
talk given there closely followed work obtained in collaboration with 
Balasubramanian and published in~\cite{bl98}. This paper reports on new 
results.} 
}

\author{Finn Larsen}

\address{Department of Physics and Astronomy\\
University of Pennsylvania\\
Philadelphia, PA 19104}


\maketitle\abstracts{At low energy the near horizon geometry of
nonextreme black holes in four dimensions exhibits an effective 
$SL(2,\IR)_L\times SL(2,\IR)_R$ symmetry. The parameters of the
corresponding induced conformal field theory gives the correct
expression for the black hole entropy. The resulting
spectrum of the Schwarzchild black hole is compared with another 
proposal.}

\section{Introduction}
There has been many recent attempts at deriving the entropy of
Schwarzchild black holes in four dimensions from string 
theory, but a fully satisfying understanding has not yet been 
achieved~\cite{hsm,susskind96a,horpol,martli,matrixbh,sfetsos,lowe,englert}.
This paper elaborates on the non-extreme black hole counting 
in~\cite{cy96b,fl97,cl97a,cl97c}. It follows the
spirit of the AdS/CFT correspondance~\cite{juanads,btzentropy} and
exploits the symmetries of the near-horizon region to learn about general
non-extreme black holes, including Schwarzchild black holes. 
The result provides an appealing generalization of the
computations for the dilute-gas black holes that are standard by now.
However, just as other computations of the Schwarzchild entropy in 
string theory, this work is presently without secure foundation in the 
microscopic theory, and thus speculative.

\section{The Black Hole Background}
The starting point is a large class of four-dimensional black holes  
specified by their mass $M$ and four $U(1)$ charges $Q_i$ or, more 
conveniently, parametrized by the non-extremality parameter $m$ 
and four boosts $\delta_i$~\cite{cystrings}:
\be
G_4 M = {1\over 4}m\sum_{i=0}^3 \cosh 2\delta_i~;~~~ 
G_4 Q_i = {1\over 4}m\sinh 2\delta_i~~~(i=0,1,2,3)~,
\ee
where $G_4$ is the four-dimensional Newton's constant. The metric is:
\be
ds^2_4 = -{1\over\sqrt{H_0 H_1 H_2 H_3}}(1-{2m\over r})dt^2 +
\sqrt{H_0 H_1 H_2 H_3}({1\over 1-{2m\over r}}dr^2 + r^2 d\Omega^2_2)~,
\label{eq:metric}
\ee
where $H_i = 1 + 2m\sinh^2\delta_i /r$. The Reissner-Nordstr\"{o}m black
hole corresponds to the case where the four $U(1)$ charges are identical 
$Q_0=Q_1=Q_2=Q_3\equiv {1\over 4}Q_{\rm RN}$; the Schwarzchild black
hole is the special case $Q_{\rm RN}=0$. From the metric 
eq.~\ref{eq:metric} one finds the black hole entropy:
\be
S\equiv {A_4\over 4G_4} = {4\pi m^2\over G_4}~\prod^3_{i=0}\cosh\delta_i~,
\label{eq:macroent}
\ee
where $A_4$ is the area of the outer horizon.

The dilute gas black holes satisfy $\delta_{1,2,3}\gg 1$, and their 
near horizon geometry is ${\rm BTZ}\times S^2$. The BTZ black hole 
is locally $AdS_3$~\cite{btz}; so in this case the CFT/AdS
correspondance applies directly~\cite{btzentropy,bl98}. 
However, the dilute gas conditions imply a near extreme limit that does 
not include the Reissner-Nordstr\"{o}m black holes. In this
paper the CFT/AdS correspondance is applied without any conditions on 
the black hole parameters.

\section{The Wave Equation}
The black hole geometry can be analyzed by considering test fields 
that propagate in the background, {\it e.g.}, a minimally 
coupled scalar field satisfying the massless Klein-Gordon equation:
\be
{1\over\sqrt{-g}}\partial_\mu (\sqrt{-g}g^{\mu\nu}\partial_\nu \Phi) =0~.
\ee
The variables are separated by writing the wave function in spherical 
coordinates as:
\be
\Phi\equiv\Phi_r(r)~\chi(\theta)
e^{-i\omega t+im\phi}~,
\ee
and we introduce the dimensionless radial 
coordinate:
\be
x \equiv {r - {1\over 2}(r_{+}+r_{-})\over r_{+}-r_{-}}~,
\label{eq:xdef}
\ee
where $r_+ = 2m$ and $r_- = 0$. Then the outer and inner event horizons 
are located at $x=\pm {1\over 2}$, respectively; and the asymptotic
space is at large $x$. The radial wave equation becomes:
\be
{\partial\over\partial x}(x^2-{1\over 4}){\partial\over\partial x}\Phi_r
+[{1\over x-{1\over 2}}
{\omega^2\over 4\kappa^2_{+}}
-{1\over x+{1\over 2}}{\omega^2\over 4\kappa_{-}^2} + V(x)]\Phi_r = 
l(l+1)\Phi_r~, 
\label{eq:geneq}
\ee
where the surface accelerations $\kappa_\pm$ at the outer and
inner horizons are:
\be
{1\over\kappa_+} = 4m\prod_i \cosh\delta_i~~~;~~
{1\over\kappa_-} = 4m\prod_i \sinh\delta_i~,
\ee
respectively; and the effective potential is:
\be
V(x) = 4x^2 m^2\omega^2 + x~8G_4 M m\omega^2 + 
(1+\sum_{i<j}\cosh 2\delta_i\cosh 2\delta_j)m^2 \omega^2 ~.
\ee
The radial wave equation has pole terms 
at the horizons $x=\pm {1\over 2}$ 
that are characteristic of black holes; they encode the 
distinctive features of the causal structure. In contrast, the potential
$V(x)$ contains the boundary condition that space is asymptotically
flat; and that gravitational potentials at large distances are governed 
by Newton's $1/r$ law. Thus the potential $V(x)$ contains the
``unimportant'' part of the black hole geometry that is unlikely to 
reveal much about its internal structure.
 
There are several circumstances where it can be justified to neglect the
potential $V(x)$. Here we exploit that {\it the potential $V(x)$ is 
neglible when the frequency of the probe is small}. In contrast, the 
standard strategy for computations of greybody factors in string theory
imposes conditions on both the black hole 
parameters and the probe energy~\cite{greybody}.
The precise low-energy condition $M\omega\ll 1$ needed here ensures the 
universal low-energy cross-section 
$\sigma_{\rm abs}(\omega\rightarrow 0)=A_4$ as well; so it is 
reasonable to expect that this condition also suffices to reveal the 
model-independent aspects of black hole entropy.

The radial wave equation exhibits an interesting structure when the 
potential $V(x)$ is omitted: it realizes
the group $SL(2,\IR)_L\times SL(2,\IR)_R$ that is also the world-sheet 
conformal symmetry group of string theory~\cite{cl97a}. This
feature of the background geometry can be exploited to shed light
on the underlying effective string theory.
According to the previous paragraph the $SL(2,\IR)_L\times SL(2,\IR)_R$ 
symmetry applies to {\it all} black holes at very low energy.
 
It is customary to interpret the four dimensional dilute gas black 
hole as a five dimensional black string. This effective string in 
turn arises as the intersection of three M5-branes that are mutually 
orthogonal and wrapped on a small six-torus~\cite{bl96,kt}. From the 
five dimensional point of view the near horizon geometry of the 
M5-branes is $AdS_3\times S^2$
and the isometries of the $AdS_3$ accounts for the 
$SL(2,\IR)_L\times SL(2,\IR)_R$ symmetry~\cite{skenderis97a,juanads}. 
However, this interpretation is not mandatory. Probes that 
are massless in four dimensions are independent of the fifth direction 
so, for such probes, the ``extra'' coordinate is a {\it redundant} 
variable whose role is to realize the $SL(2,\IR)_L\times SL(2,\IR)_R$ 
linearly~\cite{cl97a}. If we interpret the $SL(2,\IR)_L\times
SL(2,\IR)_R$ symmetry of the near-horizon geometry in this way it may 
persist even when the higher dimensional interpretation has no
preferred direction which, together with the $t$ and $r$ coordinates, 
could form the $AdS_3$. From this abstract point of view, the symmetry makes 
sense for black holes that do not satisfy the dilute gas condition.

\section{Counting Black Hole Microstates}
The working hypothesis is that the underlying microscopic theory is 
similar to the one governing the BTZ black hole; {\it i.e.} a two
dimensional conformal field theory with the central charges~\cite{adsc}:
\be
c_L = c_R  = {3\lambda\over 2G_3}~,
\label{eq:ceff}
\ee
where the effective cosmological constant in three dimensions
is $\Lambda= -\lambda^2$. 

The wave equation for a probe in the BTZ 
background~\cite{sachs97} agrees precisely with eq.~\ref{eq:geneq},
for $V(x)=0$, if we make the identifications:
\be
\beta_{L,R} \equiv {2\pi\over\kappa_+} \mp {2\pi\over\kappa_-}
= 8\pi m (\prod_{i=0}^3\cosh\delta_i \mp \prod_{i=0}^3\sinh\delta_i )
= {2\pi\lambda R_{11}\over\sqrt{M_3\lambda^2 \mp 8\lambda G_3 J_3}}~,
\label{eq:betaeff}
\ee
where $M_3$ and $J_3$ are the parameters of the BTZ black hole and
the auxiliary length $R_{11}$ relates the four dimensional Schwarzchild 
time and the BTZ-time through $\lambda t_4=t_{\rm BTZ}R_{11}$. Then the 
conformal weights of the underlying quantum states: 
\be
h_{L,R} = {M_3\lambda^2 \mp 8\lambda G_3 J_3\over 16\lambda G_3}~,
\ee
determine the entropy of the left- and right-movers as:
\be
S_{L,R}=
2\pi\sqrt{{c_{L,R}\over 6}h_{L,R}} = {\lambda\over 
32 G_3 m (\prod_{i=0}^3\cosh\delta_i \mp \prod_{i=0}^3\sinh\delta_i )}~
2\pi R_{11}.
\label{eq:slr}
\ee
Alternatively, this result follows from the standard expression for
entropy of gasses in one spatial dimension of length 
$L=2\pi R_{11}$:
\be
S_{L,R} = {c_{L,R}\over 6}~{\pi\over\beta_{L,R}}~L~,
\label{eq:gasent}
\ee
with the parameters given in eqs.~\ref{eq:ceff} 
and~\ref{eq:betaeff}~\cite{kastor}.

A low energy wave that is incident on the four-dimensional black hole
experiences a ${\rm BTZ}\times S^2$ geometry. It has no dependence
on the angular BTZ variable, the ``extra coordinate'', and the radius of
the sphere is related to the cosmological constant as 
$R_{\rm sph} = {1\over 2}\lambda$, by the equations of motion. Comparing
the dimensional reductions of this five-dimensional geometry to four
physical dimensions, and to three BTZ-dimensions, we find~\cite{bl98}:
\be
{1\over G_3} ={1\over G_4}~{\lambda^2\over 2R_{11}}~.
\label{eq:g3}
\ee
Thus the $S_{L,R}$ in eq.~\ref{eq:slr} are independent of $R_{11}$, as they 
should be.

The thermodynamic argument that leads to eq.~\ref{eq:gasent} for
free gasses implies that the $N_{L,R}\equiv {c_{L,R}\over 6} h_{L,R}$
are quantized with integer spacings. On the other hand, quantization
rules on the string theory charges give the relation 
$64 G^2_4 \prod_{i=0}^3 Q_i = \prod_{i=0}^3 n_ i = {\rm integer}$,
for general non-extreme black holes~\cite{fl97}. Consistency then
determines the cosmological constant as:
\be
\lambda= 4m (\prod_{i=0}^3\cosh^2\delta-\prod_{i=0}^3
\sinh^2\delta)^{1\over 3}~,
\label{eq:lam}
\ee
after a short computation of $N_L-N_R$. 
This generalizes the dilute gas result 
$\lambda =8G_4 (Q_1 Q_2 Q_3)^{1\over 3}$, in a way that 
treats all four charges symmetrically. 

We now find:
\be
S = 2\pi ( \sqrt{{c_L\over 6} h_L } + \sqrt{{c_R\over 6}h_R})
 = {4\pi m^2\over G_4} \prod_{i=0}^3\cosh\delta_i ~,
\ee 
in perfect agreement with eq.~\ref{eq:macroent}. In particular, 
the full functional dependence on all charges is reproduced correctly.

Angular momentum was not included in the above. However, this can
be easily remedied and the ensuing computation is much tighter,
giving supporting evidence for the argument. The general 
semiclassical quantization rule now states that:
\bea
N_{L} &=& {c_L\over 6} h_{L}= [{m^2\over G_4}
(\prod_{i=0}^3\cosh\delta_i + \prod_{i=0}^3\sinh\delta_i)]^2~,
\label{eq:quant1}\\
N_{R} &=& {c_R\over 6} h_{R}= [{m^2\over G_4}
(\prod_{i=0}^3\cosh\delta_i - \prod_{i=0}^3\sinh\delta_i)]^2-J^2~,
\label{eq:quant2}
\eea
have integer spacings.

The above presentation relied on the computation of BTZ
entropy by Strominger~\cite{btzentropy}. Albeit elegant, 
this calculation is not above criticism, as recently emphasized
in~\cite{carlip98}. Thus the ultimate justification for entropy 
calculations, in the dilute gas regime, remains our understanding of 
$D$-branes. Similarly, the non-extreme entropy must eventually be 
derived from the complete string theory spectrum, far from BPS. 
The form of the non-extreme black hole entropy suggests that, far from 
being complicated, this spectrum is structured in the same way as the 
near-extreme one.

As it stands, the computation of non-extreme entropy has several 
weaknesses, beyond those present in the dilute gas case. In particular, 
the ${\rm BTZ}\times S^2$ geometry is an auxiliary concept that is not 
realized explicitly; the area $\pi\lambda^2$ of the effective sphere is 
in general unrelated to the area of the outer horizon. On the other
hand, this also ascertains that the final result was not ``put in'' by
hand. Another concern is that the low energy condition $M\omega\ll 1$
imposed on the test field implies $\beta_{L,R}\omega\ll 1$, whereas
the modes responsible for the black hole entropy appears to satisfy 
$\beta_{L,R}\omega\sim 1$. This mismatch of scales is related 
to the absence of a ``decoupling limit'' ensuring that the worldvolume 
field theory does not couple to gravity. A possible interpretation is 
that near-extreme and non-extreme black holes are separated by a phase 
transition, and {\it the present work assumes that this does not 
happen}.

\section{Comparison with Area Quantization}

It is instructive to compare the quantization rule given in 
eqs.~\ref{eq:quant1}-\ref{eq:quant2} 
with the ``area quantization'' of Bekenstein-Mazur-Mukhanov 
(BMM)~\cite{bek95,bek97,mazur97} (see also~\cite{ashtekar}). Consider 
for definiteness the Kerr-Newman entropy:
\be
S = 2\pi [ (G_4 M^2 - {1\over 2}G_4 Q^2_{\rm KN})
+ \sqrt {G^2_4 M^4 - G^2_4 M^2 Q^2_{\rm KN}-J^2}]~.
\ee
Eqs.~\ref{eq:quant1}-\ref{eq:quant2} imply that the expression under the
square root and the square of the expression in round brackets are both 
quantized with integer spacings. The BMM prescription states that only
the square bracket is so quantized. In particular, 
eqs.~\ref{eq:quant1}-\ref{eq:quant2} gives $G^2_4 M^4={\rm integer}$ for 
Schwarzchild black holes; in contrast, BMM advocate 
$G_4 M^2={\rm integer}$.

These differences can be elucidated by considering various extreme rotating
black holes. For neutral Kerr black holes the expression under the 
square root vanishes in the extreme limit, and so $G_4 M^2=J={\rm
integer}$. If this is the complete spectrum also in the Schwarzchild
case, we find the BMM result. However, in the dilute gas limit
we trust eqs.~\ref{eq:quant1}-\ref{eq:quant2}; in particular:
\be
N_L - N_R = q + J^2~,
\ee
where the quantized charge $q$ has integer spacings. In the extreme
limit $N_R\rightarrow 0$; so, for {\it exactly neutral black holes} 
($q=0$), the quantum number $N_L$ is the {\it square} of an integer.
Since the extremal entropy $S=2\pi\sqrt{N_L}$, this is again the BMM 
result. However, for {\it generic (charged) black holes} the $N_L$ 
can be {\it any} integer, even in the extreme limit; so the allowed
masses are generically spaced much closer than the BMM argument
indicates. If this is the complete spectrum also in the Schwarzchild
case, we find the quantization rule advocated here.

\section*{Acknowledgments}
I thank V. Balasubramanian and M. Cveti\v{c} for collaborations that 
contributed to the work presented here; and the Physics Department
of the University of Michigan in Ann Arbor for hospitality while
the manuscript was prepared. This work was supported in part by DOE 
grant DOE-FG02-95ER40893.


\begin{thebibliography}{10}

\bibitem{bl98}
V.~Balasubramanian and F.~Larsen.
\newblock Near horizon geometry and black holes in four dimensions.
\newblock hep-th/9802198.

\bibitem{hsm}
G.~Horowitz, J.~Maldacena, and A.~Strominger.
\newblock Nonextremal black hole microstates and {U} duality.
\newblock {\em Phys.Lett.B}, 38:151--159, 1996.
\newblock hep-th/9603109.

\bibitem{susskind96a}
E.~Halyo, B.~Kol, A.~Rajaraman, and L.~Susskind.
\newblock Counting {S}chwarzchild and charged black holes.
\newblock {\em Phys.Lett.B}, 401:15--20, 1997.
\newblock hep-th/9609075.

\bibitem{horpol}
Horowitz and J.~Polchinski.
\newblock A correspondence principle for black holes and strings.
\newblock {\em Phys.Rev.D}, 55:6189--619, 1997.
\newblock hep-th/9612146.

\bibitem{martli}
M.~Li and E.~Martinec.
\newblock On the entropy of matrix black holes.
\newblock {\em Class.Quant.Grav.}, 14:3205--3213, 1997.
\newblock hep-th/9704134.

\bibitem{matrixbh}
T.~Banks, W.~Fischler, I.R. Klebanov, and L.~Susskind.
\newblock Schwarzschild black holes from matrix theory.
\newblock {\em Phys.Rev.Lett}, 80:226--229, 1998.
\newblock hep-th/9709091.

\bibitem{sfetsos}
K.~Sfetsos and K.~Skenderis.
\newblock Microscopic derivation of the {B}ekenstein-{H}awking entropy formula
  for nonextremal black holes.
\newblock {\em Nucl.Phys.B}, 517:179--204, 1998.
\newblock hep-th/9711138.

\bibitem{lowe}
D.~Lowe.
\newblock Statistical origin of black hole entropy.
\newblock hep-th/9802173.

\bibitem{englert}
R.~Argurio, F.~Englert, and L.~Houart.
\newblock Statistical entropy of the four-dimensional {S}chwarzschild black
  hole.
\newblock hep-th/9801053.

\bibitem{cy96b}
M.~Cveti\v{c} and D.~Youm.
\newblock Entropy of non-extreme charged rotating black holes in string theory.
\newblock {\em Phys. Rev. D}, 54:2612--2620, 1996.
\newblock hep-th/9603147.

\bibitem{fl97}
F.~Larsen.
\newblock A string model of black hole microstates.
\newblock {\em Phys. Rev. D}, 56:1005--1008, 1997.
\newblock hep-th/9702153.

\bibitem{cl97a}
M.~Cveti\v{c} and F.~Larsen.
\newblock General rotating black holes in string theory: Greybody factors and
  event horizons.
\newblock {\em Phys.Rev.D}, 56:4994--5007, 1997.
\newblock hep-th/9705192.

\bibitem{cl97c}
M.~Cveti\v{c} and F.~Larsen.
\newblock Black hole horizons and the thermodynamics of strings.
\newblock {\em Nucl. Phys. B (Proc. Suppl.)}, 62:443--456, 1998.
\newblock hep-th/9706071.

\bibitem{juanads}
J.~Maldacena.
\newblock The large {N} limit of superconformal field theories and
  supergravity.
\newblock hep-th/9711200.

\bibitem{btzentropy}
A.~Strominger.
\newblock Black hole entropy from near horizon microstates.
\newblock hep-th/9712251.

\bibitem{cystrings}
M.~Cveti\v{c} and D.~Youm.
\newblock {BPS} saturated and nonextreme states in abelian {K}aluza-{K}lein
  theory and effective {N=4} supersymmetric string vacua.
\newblock In {\em STRINGS 95: Future Perspectives in String Theory}, pages
  131--147. Los Angeles, CA, 1995.
\newblock hep-th/9508058.

\bibitem{btz}
M.~Banados, Teitelboim, and J.~Zanelli.
\newblock The black hole in three-dimensional space-time.
\newblock {\em Phys.Rev.Lett.}, 69:1849, 1992.
\newblock hep-th/9204099.

\bibitem{greybody}
J.~Maldacena and A.~Strominger.
\newblock Black hole greybody factors and {D}-brane spectroscopy.
\newblock {\em Phys. Rev. D}, 55:861--870, 1996.
\newblock hep-th/9609026.

\bibitem{bl96}
V.~Balasubramanian and F.~Larsen.
\newblock On {D}-branes and black holes in four dimensions.
\newblock {\em Nucl. Phys.B}, 478:199, 1996.
\newblock hep-th/9604189.

\bibitem{kt}
I.~Klebanov and A.~Tseytlin.
\newblock Intersecting {M}-branes as four-dimensional black holes.
\newblock {\em Nucl. Phys.B}, 475:179--192, 1996.
\newblock hep-th/9604166.

\bibitem{skenderis97a}
H.~Boonstra, B.~Peeters, and K.~Skenderis.
\newblock Duality and asymptotic geometries.
\newblock {\em Phys.Lett.B}, 411:59--67, 1997.
\newblock hep-th/9706192.

\bibitem{adsc}
J.D. Brown and M.~Henneaux.
\newblock Central charges in the canonical realization of asymptotic
  symmetries: An example from three-dimensional gravity.
\newblock {\em Comm.Math.Phys}, 104:207--226, 1986.

\bibitem{sachs97}
D.~Birmingham, I.~Sachs, and S.~Sen.
\newblock Three-dimensional black holes and string theory.
\newblock {\em Phys.Lett.B}, 413:281--286, 1997.
\newblock hep-th/9707188.

\bibitem{kastor}
D.~Kastor and J.~Traschen.
\newblock A very effective string model?
\newblock {\em Phys.Rev.D}, 57:4862--4869, 1998.
\newblock hep-th/9707157.

\bibitem{carlip98}
S.~Carlip.
\newblock What we don't know about {BTZ} black hole entropy.
\newblock hep-th/9806026.

\bibitem{bek95}
J.~Bekenstein and S.~Mukhanov.
\newblock Spectroscopy of the quantum black hole.
\newblock {\em Phys.Lett.B}, 360:7--12, 1995.
\newblock gr-qc/9505012.

\bibitem{bek97}
J.~Bekenstein.
\newblock Quantum black holes as atoms.
\newblock gr-qc/9710076.

\bibitem{mazur97}
P.~Mazur.
\newblock On gravitation and quanta.
\newblock hep-th/9712208.

\bibitem{ashtekar}
A.~Ashtekar, J.~Baez, A.~Corichi, and K.~Krasnov.
\newblock Quantum geometry and black hole entropy.
\newblock {\em Phys.Rev.Lett}, 80:904--907, 1998.
\newblock gr-qc/9710007.

\end{thebibliography}

\end{document}